\documentclass[twocolumn,preprintnumbers,amsmath,amssymb]{revtex4}
\usepackage{graphicx}% Include figure files
\usepackage{dcolumn}% Align table columns on decimal point
\usepackage{bm}% bold math
\usepackage{textcomp}

\begin{document}

\title{Protecting quantum states from decoherence of finite temperature using weak measurement}

\author{Shu-Chao Wang$ ^{1}$, Zong-Wen Yu$ ^{2}$,
and Xiang-Bin Wang$ ^{1,3\footnote{Email
Address:xbwang@mail.tsinghua.edu.cn}}$}

\affiliation{ \centerline{$^{1}$State Key Laboratory of Low
Dimensional Quantum Physics, Tsinghua University, Beijing 100084,
People¡¯s Republic of China}\centerline{$^{2}$Data Communication Science and Technology Research Institute, Beijing 100191, China}\centerline{$^{3}$ Shandong
Academy of Information and Communication Technology, Jinan 250101,
People¡¯s Republic of China}}

\date{\today}

\begin{abstract}
We show how to optimally protect quantum states and quantum entanglement under non-zero temperature based on measurement reversal from weak measurement. In particular, we present explicit formulas of the protection.
\end{abstract}

\maketitle
\section{Introduction}
\parskip=0 pt
The inherit properties of quantum mechanics can be applied to nontrivial tasks in quantum information processing(QIP) such as the design of fast computation, the unconditionally secure private communication. However, in practice, the decoherence can undermine severly the quantum features in QIP. Protecting quantun states and quantum entanglement under decoherence is crucially important in effective QIP. Many proposals have been suggested for quantum coherence protection including passive methods,e.g,decoherence-free\cite{free1,free2,free3} subspaces and active methods like quantum error correction code\cite{qecc1,qecc2,qecc3} , the technique of dynamical decoupling \cite{dd1,dd2,dd3}or using quantum Zeno dynamics\cite{zeno1,zeno2}. When the decoherence is due to processes with short correlation time scales, it is shown that quantum reversal scheme has advantages\cite{pra,nature,enhance1,enhance2}.Weak measurements has also been found useful in entanglement amplification\cite{amplification}.Quantun entanglement plays an essential role in quantum information processing and gives rise to varieties of interesting phenomena\cite{nielsen}. But it is fragile to environmental noises. It is of great meaning to protect quantum entanglement.

Recently, a novel idea\cite{pra,nature} is proposed to protect quantum states and quantum entanglements from decoherence using weak measurement and measurement reversal. However, their result is limited to a special class of channel noise, which corresponds to the zero temperature environmental noise. Most often, decoherence is caused by the uncontrollable interaction with the environment. In the case of zero temperature, a type of noise due to environmental interaction can be modeled as the following  amplitude damping(AD) channel\cite{nielsen}:
\begin{equation}
{\varepsilon _{AD}}(\rho ) = \sum\limits_{i = 0}^1 {{E_i}\rho E_i^\dag }
\end{equation}
with
\begin{equation}
{E_0} = \left( {\begin{array}{*{20}{c}}
1&0\\
0&{\sqrt {1 - r} }
\end{array}} \right),{E_1} = \left( {\begin{array}{*{20}{c}}
0&{\sqrt r }\\
0&0
\end{array}} \right)\label{eqad}
\end{equation}
 It has been shown\cite{pra,nature} that quantum state and quantum entanglement can be effectively protected under such a channel. However, in practice, environmental temperature cannot be zero. In non-zero temperature , the channel is more complicated than Eq.(\ref{eqad}). An important class of dissipation under finite temperature can be modeled by the following generalized amplitude (GAD) channel\cite{nielsen}:
 \begin{equation}
 {\varepsilon _{GAD}}(\rho ) = \sum\limits_{i = 0}^3 {{E_i}\rho E_i^\dag }
 \end{equation}
 with
 \begin{equation}
\begin{array}{l}
{E_0} = \sqrt p \left( {\begin{array}{*{20}{l}}
1&0\\
0&{\sqrt {1 - r} }
\end{array}} \right),{E_1} = \sqrt p \left( {\begin{array}{*{20}{l}}
0&{\sqrt r }\\
0&0
\end{array}} \right)\\
{E_2} = \sqrt {1 - p} \left( {\begin{array}{*{20}{l}}
{\sqrt {1 - r} }&0\\
0&1
\end{array}} \right),{E_3} = \sqrt {1 - p} \left( {\begin{array}{*{20}{l}}
0&0\\
{\sqrt r }&0
\end{array}} \right)
\end{array}.\label{eqgad}
\end{equation}

One can see that Eq.(\ref{eqgad}) reduces to Eq.(\ref{eqgad})when $p=1$. In the GAD channel, an atom can not only transit from the higher energy level to the lower one by undergoing spontaneous emission, but also can jump from the lower energy state to the higher energy state by absorbing energy from the finite-temperature environment. Generalized amplitude damping describes the finite-temperature relaxation processes due to coupling of spins to their surrounding lattice, a large system which is in thermal equilibrium at a temperature often much higher than the spin temperature\cite{nielsen}.

 In this work, we study how to use weak measurement to battle against the decoherence in such channels. By performing weak measurements and measurement reversals, the final fidelity can be optimized by adjusting the measurement parameters. We have also investigated how to use weak measurements to recover quantum entanglement at finite temperature environment. Explicit formulas for optimal results are presented.

This article is organized as follows. In the following section, we show how to use weak measurement to protect qubit states against decoherence in generalized amplitude channel. The average fidelity over the initial state is also studied. The optimal measurement strength is given. In the third section, we study how to use weak measurements to protect quantum entanglement in GAD channels, we present an optimal measurement strength for obtaining most entanglement. The article is ended with a concluding remark.

\section{protect quantum qubit through weak measurements}

Any pure qubit state can be written as a vector on the Bloch-sphere: $\rho  = \frac{1}{2}(I + \sin \theta \cos \varphi X + \sin \theta \sin \varphi Y + \cos \theta Z)$. Let us first consider the equatorial states ($\theta  = \frac{\pi }{2}$) which are extensively applied in QKD\cite{qkd}. In this case, the initial state can be written as
\begin{equation}
{\rho _{in}} = \frac{1}{2}\left( {\begin{array}{*{20}{c}}
1&{{e^{ - i\varphi }}}\\
{{e^{i\varphi }}}&1
\end{array}} \right).
\end{equation}

Under the GAD channel as as described by Eq.(\ref{eqgad}), due to decoherence, the outcome state is a mixed state,
\begin{equation}
{\rho _f} = \frac{1}{2}\left( {\begin{array}{*{20}{c}}
{1 - r + 2pr}&{\sqrt {1 - r} {e^{-i\varphi }}}\\
{\sqrt {1 - r} {e^{  i\varphi }}}&{1 + r - 2pr}
\end{array}} \right)
\end{equation}
The fidelity of the initial state and this state is,
\begin{equation}
F = \frac{1}{2}(1 + \sqrt {1 - r} )
\end{equation}

In order to improve the fidelity, we should perform two weak measurements $M$ and $N$, before and after the qubit being put into the GAD channel,respectively. With these weak measurements being implemented, the final state is,
\begin{equation}
{\rho _{f}^{(w)}} = N{\varepsilon _{GAD}}(M{\rho _{in}}{M^\dag }){N^\dag }
\end{equation}
with $\epsilon_{GAD}$ being defined by Eq.(3) and the non-unitary quantum operations
 \begin{equation}
 M = \left( {\begin{array}{*{20}{c}}
1&0\\
0&m
\end{array}} \right)
\end{equation}
and
\begin{equation}
N = \left( {\begin{array}{*{20}{c}}
n&0\\
0&1
\end{array}} \right).
\end{equation}
It's easy to see,
\begin{equation}
{\rho _{f}^{(w)}} = \frac{1}{T}\left( {\begin{array}{*{20}{c}}
{{n^2}(pr{m^2} + pr - r + 1)}&{mn\sqrt {1 - r} {e^{ - i\varphi }}}\\
{mn\sqrt {1 - r} {e^{i\varphi }}}&{ - pr{m^2} + {m^2} - pr + r}
\end{array}} \right),
\end{equation}
 and
 \begin{equation}
 T = {n^2}(pr{m^2} + pr - r + 1) - pr{m^2} + {m^2} - pr + r
 \end{equation}
  is the normalization factor. The fidelity of the final state and the initial state is
 \begin{equation}
 F^{(w)} = \frac{1}{2} + \frac{{mn\sqrt {1 - r} }}{{{n^2}(pr{m^2} + pr - r + 1) + {m^2}(1 - pr) + (1 - p)r}}.
\end{equation}
The overall success probability is
\begin{equation}
{P_s} = \frac{T}{2}\cdot\min \left\{ {1,\frac{1}{m}} \right\}\cdot\min \left\{ {1,\frac{1}{n}} \right\}.
\end{equation}
When
\begin{equation}
m = \sqrt[4]{{\frac{{(1 - p)(1 - r + pr)}}{{p(1 - pr)}}}},n = \sqrt[4]{{\frac{{(1 - p)(1 - pr)}}{{p(1 - r + pr)}}}}\label{cond},
\end{equation}
we obtain the maximal value for $F^{(w)}$ as
\begin{equation}
{F^{(w)}_{\max }} = \frac{1}{2}\left( {1 + \frac{{\sqrt {1 - r} }}{{\sqrt {(1 - pr)(1 - r + pr)}  + r\sqrt {p(1 - p)} }}} \right).
\end{equation}

\begin{figure}
   %Requires \usepackage{graphicx}
  \includegraphics[width=240pt]{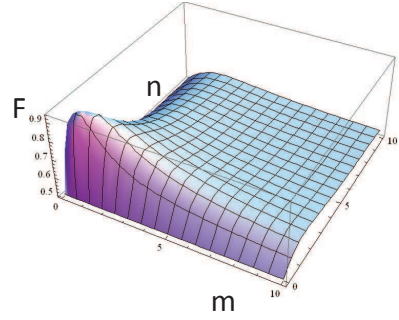}\\
  \caption{The fidelity with varying measurement strength m and n with p=0.8 and r=0.3. One can find that the optimal value of m and n is zero as in the amplitude channel (p=1) case. But in general when $p\neq1$,the optimal value of m and n are not zero.}\label{fid}
\end{figure}

One can check that ${F_{\max }} \ge {F_0}$(see Appendix A), so weak measurement is useful when we consider the equatorial states, under the generalized amplitude damping channel. By adjusting the measurement strengths according to the channel parameters, one can obtain a final state with a higher fidelity.

In the quantum key distribution, the equatorial states can be used as BB84 states\cite{nielsen}. The two basis $\left\{ {\left| 0 \right\rangle  \equiv \frac{1}{{\sqrt 2 }}\left( {\left| 0 \right\rangle  + \left| 1 \right\rangle } \right),\left| 1 \right\rangle  \equiv \frac{1}{{\sqrt 2 }}\left( {\left| 0 \right\rangle  - \left| 1 \right\rangle } \right)} \right\}$ and $\left\{ {\left| 0 \right\rangle  \equiv \frac{1}{{\sqrt 2 }}\left( {\left| 0 \right\rangle  + i\left| 1 \right\rangle } \right),\left| 1 \right\rangle  \equiv \frac{1}{{\sqrt 2 }}\left( {\left| 0 \right\rangle  - i\left| 1 \right\rangle } \right)} \right\}$ can be used to complete the quantum key distribute processing. The error rate can be defined as
\begin{equation}
{R_E} = {\left\langle {\frac{{\left\langle i \right|{\rho _{i \oplus 1}}\left| i \right\rangle }}{{\left\langle i \right|{\rho _i}\left| i \right\rangle  + \left\langle i \right|{\rho _{i \oplus 1}}\left| i \right\rangle }}} \right\rangle _i}.
\end{equation}
Here, $\rho_i$ means the obtained density matrix of the qubit after undergoing the weak measurements and the GAD channel when the initial state is $|i\rangle$, i=0 or 1 and ${\left\langle  \bullet  \right\rangle _i}$ denotes the average over the 4 basis states $\frac{1}{{\sqrt 2 }}\left( {\left| 0 \right\rangle  \pm (i)\left| 1 \right\rangle } \right)$. By calculating, one can find that ${R_E} = 1 - {F^{(w)}}$, which means that while maximizing the fidelity, we also minimize the error rate.

We can also maximize the averaged fidelity $\bar{F}$ over six symmetric states on the Bloch sphere. For experimentally
characterizing quantum gates and channels, it is meaningful to consider the average fidelity $\overline F$ over only six initial states\cite{pra,six}:$|0\rangle,|1\rangle,(|0\rangle\pm|1\rangle)/\sqrt{2},(|0\rangle\pm i|1\rangle)/\sqrt{2}$. Without any weak measurement, one can calculate that, the final fidelity after passing the channel is $F_0=1-r+pr$ for the initial state $|0\rangle$; $F_1=1-pr$, for $|1\rangle$; $F_e=\frac{1}{2}(1+\sqrt{1-r})$, for the equatorial states $(|0\rangle\pm|1\rangle)/\sqrt{2}$ and $(|0\rangle\pm i|1\rangle)/\sqrt{2}$. So we have,
\begin{equation}
\overline F  = \frac{1}{6}({F_0} + {F_1} + 4{F_e}) = \frac{1}{3} + \frac{1}{6}{(1 + \sqrt {1 - r} )^2}.
\end{equation}

With the weak measurements $M$ and $N$ given above, we can get the average fidelity of these six states,
\begin{widetext}
\begin{equation}
{\overline F ^{(w)}} = \frac{1}{3} + \frac{1}{6}\left( {\frac{{{n^2}(1 - r + rp)}}{{r - rp + {n^2}(1 - r + rp)}} + \frac{{1 - rp}}{{1 - rp + {n^2}rp}} + \frac{{4mn\sqrt {1 - r} }}{{{n^2}(pr{m^2} + pr - r + 1) + {m^2}(1 - pr) + (1 - p)r}}} \right).
\end{equation}
\end{widetext}
We can show that when Eq.(\ref{cond}) is satisfied, ${\overline F ^{(w)}}$ has the maximal value (see Appendix B). Note that when $p=1$, then $m \to 0,n \to 0$, and the optimal measurements becomes projective measurements. This coincides with the the previous results\cite{pra}.

\section{protect entanglement through weak measurements}

 \parskip=0 pt
 Quantum entanglement plays an important role in the quantum information processing. But it is very fragile due to the decoherence.  We now study how the GAD channel affects a two-qubit entangled state. The channel can also be described as the interaction of the system and the environment with the initial state ${\left| {00} \right\rangle _E}$:
 \begin{widetext}
\begin{equation}
\begin{array}{l}
{\left| 0 \right\rangle _S}{\left| {00} \right\rangle _E} \to \sqrt p {\left| 0 \right\rangle _S}{\left| {00} \right\rangle _E} + \sqrt {1 - p} \sqrt {1 - r} {\left| 0 \right\rangle _S}{\left| {01} \right\rangle _E} + \sqrt {1 - p} \sqrt r {\left| 1 \right\rangle _S}{\left| {11} \right\rangle _E}\\
{\left| 1 \right\rangle _S}{\left| {00} \right\rangle _E} \to \sqrt p \sqrt {1 - r} {\left| 1 \right\rangle _S}{\left| {00} \right\rangle _E} + \sqrt {pr} {\left| 0 \right\rangle _S}{\left| {10} \right\rangle _E} + \sqrt {1 - p} {\left| 1 \right\rangle _S}{\left| {01} \right\rangle _E}
\end{array}.
\end{equation}
\end{widetext}
For simplicity, we call the above channel parameterized by p,r as a GAD channel of $\{p,r\}$.

\begin{figure}
   %Requires \usepackage{graphicx}
  \includegraphics[width=240pt]{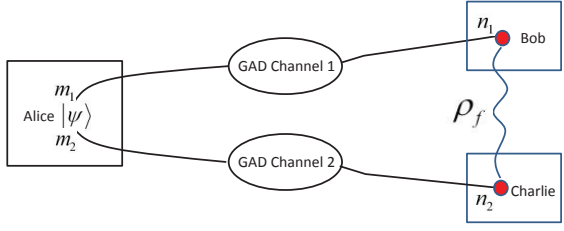}\\
  \caption{The scheme for entanglement protection using weak measurement. Initially, Alice prepare two qubits in an entangled states. Before sending the two qubits, Alice perform partial collapse weak measurement on the 2 qubits with the measurement parameter $m_1$ and $m_2$ respectively. After obtaining his qubit, Bob (Charlie) does a weak measurement with the strength $n_1$($n_2$). The concurrence can be optimized by adjusting the measurement strength. }\label{con}
\end{figure}

Suppose initially, Alice prepare the two qubits in an entangled state:
\begin{equation}
{\left| \phi  \right\rangle _{in}} = \alpha \left| {00} \right\rangle  + \beta \left| {11} \right\rangle .
\end{equation}

Then, Alice sends the two qubits to Bob and Charlie through two CAD channels characterized by $\{p_1,r_1\}$ and $\{p_2,r_2\}$. After undergoing the channels, the density matrix of the two qubits turns to be
\begin{equation}
{\rho _C} = \left( {\begin{array}{*{20}{c}}
a&0&0&e\\
0&b&0&0\\
0&0&c&0\\
{{e^*}}&0&0&d
\end{array}} \right)
\end{equation}
%\begin{widetext}
with
%\begin{subequations}\begin{align}
\begin{equation}
\begin{split}
a & = [1 - {r_1} - {r_2} + {r_1}{r_2} + {p_1}{r_1} + {p_2}{r_2} - ({p_1} + {p_2}){r_1}{r_2} \\&+ {p_1}{p_2}{r_1}{r_2}]{\left| \alpha  \right|^2} + {p_1}{p_2}{r_1}{r_2}{\left| \beta  \right|^2},  \\
b & = [{r_2} - {r_2}{p_2} - {r_1}{r_2} + ({p_1} + {p_2}){r_1}{r_2} - {p_1}{p_2}{r_1}{r_2}]{\left| \alpha  \right|^2} \\&+ ({p_1}{r_1} - {p_1}{p_2}{r_1}{r_2}){\left| \beta  \right|^2}, \\
c & = [{r_1} - {p_1}{r_1} - {r_1}{r_2} + ({p_1} + {p_2}){r_1}{r_2} - {p_1}{p_2}{r_1}{r_2}]{\left| \alpha  \right|^2}\\& + ({p_2}{r_2} - {p_1}{p_2}{r_1}{r_2}){\left| \beta  \right|^2}, \\
d & = (1 - {p_1})(1 - {p_2}){r_1}{r_2}{\left| \alpha  \right|^2} \\&+ (1 - {p_1}{r_1} - {p_2}{r_2} + {p_1}{p_2}{r_2}{r_2}){\left| \beta  \right|^2}, \\
e & = \alpha {\beta ^*}\sqrt {1 - {r_1}} \sqrt {1 - {r_2}}
%\end{align}
%\end{subequations}
\end{split}
\end{equation}
%\end{widetext}

The concurrence\cite{concorrence} of $\rho_C$ is
\begin{equation}
\mathcal{C}({\rho _C}) = \max \left\{ {0,{\Lambda _1} \equiv 2(\left| e \right| - \sqrt {bc} )} \right\}.
\end{equation}
When $\Lambda_1>0$, the concurrence is $\Lambda_1$, otherwise, the concurrence is zero.
To improve the entanglement Bob and Charlie shared, Alice chooses weak measurements on both qubits before sending them through the channel. The two-qubit weak measurement is a non-unitary operation which can be written as,
\begin{equation}
M = \left( {\begin{array}{*{20}{c}}
1&0\\
0&{{m_1}}
\end{array}} \right) \otimes \left( {\begin{array}{*{20}{c}}
1&0\\
0&{{m_2}}
\end{array}} \right)
\end{equation}
After obtaining the two qubits, Bob and Charlie does a weak measurement individually. The second weak measurement can be written as,
\begin{equation}
N = \left( {\begin{array}{*{20}{c}}
{{n_1}}&0\\
0&0
\end{array}} \right) \otimes \left( {\begin{array}{*{20}{c}}
{{n_2}}&0\\
0&0
\end{array}} \right).
\end{equation}
The final density matrix of the two qubits is
\begin{equation}
{\rho _N} = \frac{1}{P}\left( {\begin{array}{*{20}{c}}
{n_1^2n_2^2A}&0&0&{{n_1}{n_2}E}\\
0&{n_1^2B}&0&0\\
0&0&{n_2^2C}&0\\
{{n_1}{n_2}{E^*}}&0&0&D
\end{array}} \right)
\end{equation}
%\begin{widetext}
with
%\begin{subequations}

\begin{equation}
\begin{split}
A &= [1 - {r_1} - {r_2} + {r_1}{r_2} + {p_1}{r_1} + {p_2}{r_2} - ({p_1} + {p_2}){r_1}{r_2} \\&+ {p_1}{p_2}{r_1}{r_2}]{\left| \alpha  \right|^2} + m_1^2m_2^2{p_1}{p_2}{r_1}{r_2}{\left| \beta  \right|^2}\\&\equiv A_{0}+A_{1}m_1^2m_2^2,  \\
B &= [{r_2} - {p_2}{r_2} - {r_1}{r_2} + ({p_1} + {p_2}){r_1}{r_2} - {p_1}{p_2}{r_1}{r_2}]{\left| \alpha  \right|^2} \\&+ m_1^2m_2^2({p_1}{r_1} - {p_1}{p_2}{r_1}{r_2}){\left| \beta  \right|^2},\\&\equiv B_{0}+B_{1}m_1^2m_2^2 \\
C &= [{r_1} - {p_1}{r_1} - {r_1}{r_2} + ({p_1} + {p_2}){r_1}{r_2} - {p_1}{p_2}{r_1}{r_2}]{\left| \alpha  \right|^2}\\& + m_1^2m_2^2({p_2}{r_2} - {p_1}{p_2}{r_1}{r_2}){\left| \beta  \right|^2}\\&\equiv C_{0}+C_{1}m_1^2m_2^2, \\
D &= (1 - {p_1})(1 - {p_2}){r_1}{r_2}{\left| \alpha  \right|^2} \\&+ m_1^2m_2^2(1 - {p_1}{r_1} - {p_2}{r_2} + {p_1}{p_2}{r_2}{r_2}){\left| \beta  \right|^2}\\&\equiv D_{0}+D_{1}m_1^2m_2^2, \\
E &= \alpha {\beta ^*}{m_1}{m_2}\sqrt {1 - {r_1}} \sqrt {1 - {r_2}}
\end{split}\label{eqABCDE}
\end{equation}
%\end{subequations}
%\end{widetext}
and
\begin{equation}
P = n_1^2n_2^2A + n_1^2B + n_2^2C + D.
\end{equation}
The overall success probability is
\begin{equation}
{P_s}{\rm{ = }}P\prod\limits_{{\rm{c = \{ }}{{\rm{m}}_1}{\rm{,}}{{\rm{n}}_1}{\rm{,}}{{\rm{m}}_2}{\rm{,}}{{\rm{n}}_2}{\rm{\} }}} {\min \{ 1,\frac{1}{c^{2}}\} }
\end{equation}
The corresponding concurrence is
\begin{equation}
\mathcal{C}({\rho _N}) = \max \left\{ {0,{\Lambda _2} \equiv \frac{{2{n_1}{n_2}(\left| E \right| - \sqrt {BC} )}}{{n_1^2n_2^2A + n_1^2B + n_2^2C + D}}} \right\}.
\end{equation}

We can show that $\Lambda_2$ gets its maximal value when the following conditions are met (see Appendix C),
\begin{subequations}\begin{align}
{n_1} &= \sqrt[4]{{\frac{{CD}}{{AB}}}},  \\
{n_2} &= \sqrt[4]{{\frac{{BD}}{{AC}}}}
\end{align}\label{eqn}
\end{subequations}

One can see that when the above equations are satisfied, the value of $\Lambda_2$ changes only with $m_1m_2$,we can set $m_2=1$,i.e,the weak measurement on the second qubit is not a necessary and the concurrence can be optimized by adjusting $m\equiv m_1$.
Substituting Eqs.(\ref{eqn}) into the expression of $\Lambda_2$, we get
\begin{widetext}
\begin{eqnarray}
  \Lambda_2&=& \frac{{\left| E \right| - \sqrt {BC} }}{{\sqrt {BC}  + \sqrt {AD} }} \nonumber \\
  &=&\frac{|\alpha\beta|\sqrt{(1-r_1)(1-r_2)}-\sqrt{m^2 B_1 C_1 +\frac{1}{m^2}B_0 C_0 +B_1 C_0+B_0 C_1}} {\sqrt{m^2A_1 D_1 +\frac{1}{m^2}A_0 D_0 +A_1 D_0 +A_0 D_1}+\sqrt{m^2 B_1 C_1 +\frac{1}{m^2}B_0 C_0 +B_1 C_0+B_0 C_1}}. \label{eq:Lambda2of1}
\end{eqnarray}
\end{widetext}
In order to maximize the value of $\Lambda_2$, we need the following two inequalities.
\begin{equation}
\begin{split}
   &\sqrt{m^2 B_1 C_1 +\frac{1}{m^2}B_0 C_0 +B_1 C_0+B_0 C_1}  \\
  &\geq  \sqrt{2\sqrt{B_1 C_1 B_0 C_0}+B_1 C_0 +B_0 C_1}  \\
  &=\sqrt{B_1 C_0}+\sqrt{B_0 C_1}, \label{eq:fmBC}
\end{split}
\end{equation}

the equality holds when
\begin{equation}
  m^4=\frac{B_0 C_0}{B_1 C_1}.
\end{equation}
And
\begin{equation}
\begin{split}
   &\sqrt{m^2A_1 D_1 +\frac{1}{m^2}A_0 D_0 +A_1 D_0 +A_0 D_1} \nonumber \\
  &\geq \sqrt{2\sqrt{A_1 D_1 A_0 D_0}+A_1 D_0 +A_0 D_1 }\nonumber \\
  &=\sqrt{A_1 D_0}+\sqrt{A_0 D_1}, \label{eq:fmAD}
\end{split}
\end{equation}
the equality holds when
\begin{equation}
  m^4=\frac{A_0 D_0}{A_1 D_1}.
\end{equation}
Considering the expressions of $A,B,C$ and $D$ presented in Eqs.(\ref{eqABCDE}), we can easily find out that
\begin{equation}
\begin{split}
  m &= \sqrt[4]{{\frac{{{B_0}{C_0}}}{{{B_1}{C_1}}}}} = \sqrt[4]{{\frac{{{A_0}{D_0}}}{{{A_1}{D_1}}}}} \\&= \sqrt[4]{{\frac{{(1 - {p_1})(1 - {p_2})(1 - {r_1} + {r_1}{p_1})(1 - {r_2} + {r_2}{p_2})}}{{{p_1}{p_2}(1 - {r_1}{p_1})(1 - {r_2}{p_2})}}}}\frac{{|\alpha |}}{{|\beta |}}, \label{eq:mv}
\end{split}
\end{equation}
which means that the two inequalities in Eq.(\ref{eq:fmBC}) and Eq.(\ref{eq:fmAD}) take the equality sigh with the same condition $m^4=\frac{B_0 C_0}{B_1 C_1}=\frac{A_0 D_0}{A_1 D_1}$. With the value of $m$ presented in Eq.(\ref{eq:mv}), we can obtain the maximum value of $\Lambda_2$ such that
\begin{widetext}
\begin{equation}\label{eq:Lambda2}
  \overline{\Lambda}_2=\frac{\sqrt{(1-r_1)(1-r_2)}-r_1\sqrt{p_1(1-p_1)(1-r_2 p_2)(1-r_2+r_2 p_2)}- r_2\sqrt{p_2(1-p_2)(1-r_1 p_1)(1-r_1+r_1 p_1)}}{\left(r_1\sqrt{p_1(1-p_1)}+\sqrt{(1-r_1 p_1)(1-r_1 +r_1 p_1)}\right)\left(r_2\sqrt{p_2(1-p_2)}+\sqrt{(1-r_2 p_2)(1-r_2 +r_2 p_2)}\right)}.
\end{equation}
\end{widetext}
Then we get the optimal concurrence of the output state $\rho_N$
\begin{equation}\label{eq:OptCon}
  C(\rho_N)=\max\{0,\overline{\Lambda}_2\},
\end{equation}
with $n_1,n_2$ given by Eqs.(\ref{eqn}), $m_1=m$ given in Eq.(\ref{eq:mv}), and $m_2=1$. We approach a surprising result: the maximum concurrence does not depend on the parameters $\alpha$ and $\beta$. Furthermore, under the condition in Eq.(\ref{eq:mv}), the value of $n_1$ and $n_2$ can be rewrite into
\begin{eqnarray}
  n_1&=&\sqrt[4]{\frac{(c_0+c_1 h)(d_0+d_1 h)}{(a_0+a_1 h)(b_0+b_1 h)}}, \\
  n_2&=&\sqrt[4]{\frac{(b_0+b_1 h)(d_0+d_1 h)}{(a_0+a_1 h)(c_0+c_1 h)}},
\end{eqnarray}
where $x_0|\alpha|^2=X_0, x_1|\beta|^2=X_1$, with $x=a,b,c,d$, $X=A,B,C,D$, and $h=\sqrt{\frac{b_0 c_0}{b_1 c_1}}=\sqrt{\frac{a_0 d_0}{a_1 d_1}}$. This means that $n_1$ and $n_2$ also do not depend on $\alpha$ and $\beta$. Then we can optimize the success probability $P_s$ by taking
\begin{equation}
  |\alpha|^2=\frac{1}{1+h}.
\end{equation}

When $\left| \alpha  \right| = 0$ or $\left| \beta  \right|=0$, the success probability is exactly zero, which means that one can not produce quantum entanglement only by local operations if there is no entanglement initially.

In Fig.{\ref{con}},we show how the concurrence changes with $m$. We set $p_1=0.9,r_1=0.5,p_2=0.95,r_2=0.3$. One can see that the maximal value of the concurrence is 0.53, the corresponding measurement parameter is $m=0.34$, $n_1=0.50$ and $n_2=0.44$. The success probability is about $0.06$. Without any weak measurement, the concurrence is about 0.33. The concurrence can really be enhanced in the pay of success probability. We also notice that when $p=1$,then the GAD channel reduced to a AD channel, then the optimal value  m and n tends to be zero which means the measurements become strong measurements. This coincide with the result of Ref.\cite{nature}.

We have to stress that, in GAD channel, weak measurement can help to circumvent the "entanglement sudden death". In certain conditions, $\Lambda_1$ can be smaller than zero,thus the concurrence become 0, by choosing proper weak measurement parameters m and n, $\Lambda_2$ can be made non-zero under the same conditions ,e.g,$p_1=p_2=0.7,r_1=r_2=0.61$. We note that, in such a condition, we can not get any entanglement using the previous schemes in Ref.\cite{pra,nature}.

\begin{figure}
   %Requires \usepackage{graphicx}
  \includegraphics[width=240pt]{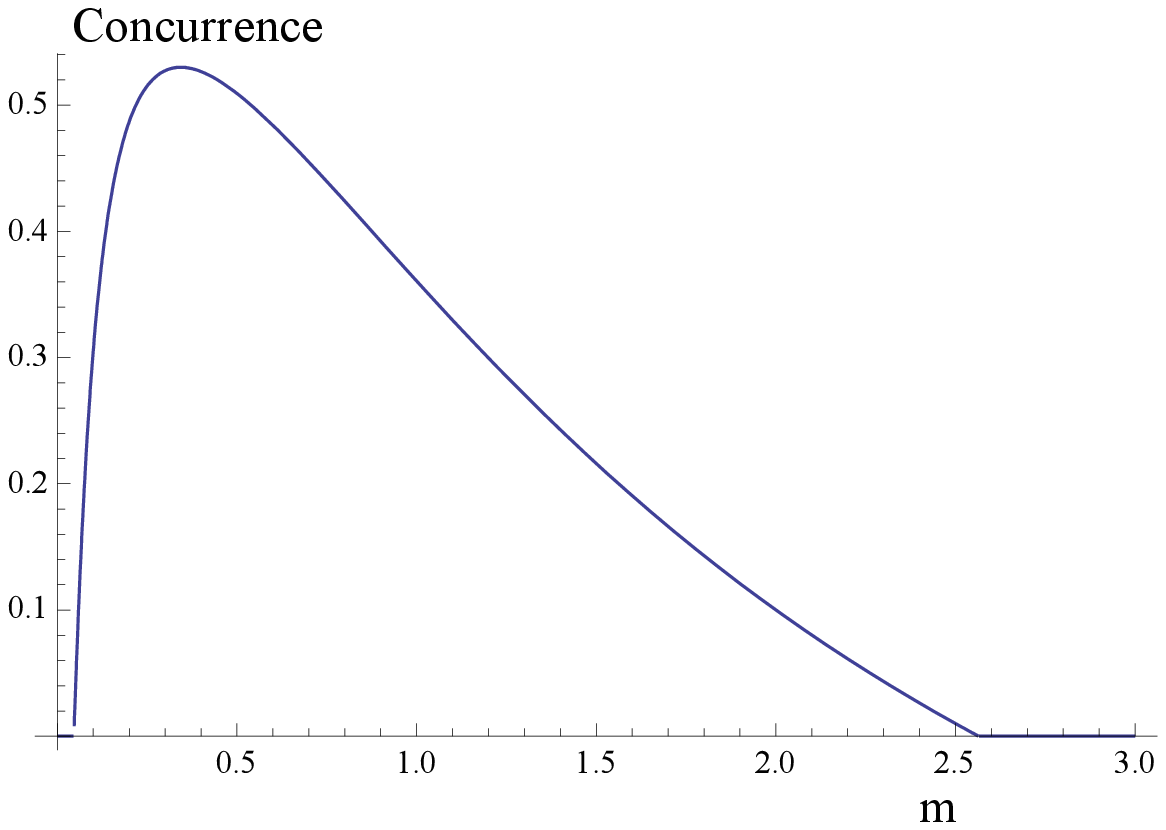}\\
  \caption{The concurrence with different m when the initial state of the two qubits is a maximally entangled state ${\left| \psi  \right\rangle _0} = \frac{1}{{\sqrt 2 }}\left( {\left| {00} \right\rangle  + \left| {11} \right\rangle } \right)$ and the channel parameters are:$p_1=0.9,r_1=0.5,p_2=0.95,r_2=0.3$. One can see that there is a optimal value for the concurrence at m=0.34. This is in agrement with Eq.(\ref{eq:mv})  }\label{con}
\end{figure}

\section{conclusion}

We have studied how weak measurement can be used for quantum state and entanglement protection exposed to environment with finite temperature. We found that the pre-channel and post-channel weak measurement are useful to battle with decoherence in generalized amplitude damping channels. For equatorial states, we give the optimal measurement strength in analysis formate. We have also shown that weak measurement are useful in protecting entanglement in finite temperature environment. When setting $p_1=1$ and $p_2=1$, our conclusion coincide with the previous results\cite{pra,nature}.

\section*{ACKNOWLEDGEMENT}
We acknowledge
the support from the 10000-Plan of Shandong province,
the National High-Tech Program of China Grants
No. 2011AA010800 and No. 2011AA010803 and NSFC
Grants No. 11174177 and No. 60725416.

\section*{APPENDIX A}

To explicate that ${F_{\max }} \ge {F_0}$, we have to study the function $G(p,r) = \sqrt {(1 - rp)(1 - r + rp)}  + r\sqrt {p(1 - p)} $. To obtain the maximal value of $G(p,r)$, we have to solve the equation:
\begin{equation}
\left\{ \begin{array}{l}
{\partial _p}G = \frac{1}{2}r(1 - 2p)\left( {\frac{1}{{\sqrt {p(1 - p)} }} + \frac{r}{{\sqrt {1 - r + (1 - p)p{r^2}} }}} \right) = 0\\
{\partial _r}G = \sqrt {p(1 - p)}  - \frac{{1 - 2pr + 2{p^2}r}}{{2\sqrt {1 - r + (1 - p)p{r^2}} }} = 0
\end{array} \right.
\end{equation}
  We can find that $G(p,r)$ has the maximal value 1 if and only if $r=0$ or $p=\frac{1}{2}$,such that ${F_{\max }} \ge {F_0}$. When $p=0$ or $p=1$, G has the minimal value $\sqrt{1-r}$, and $F_{\max }$ can be as large as 1.

\section*{APPENDIX B}
In this Appendix, we want to proof that when Eq.(\ref{cond}) is satisfied, ${\overline F ^{(w)}}$ has the maximal value. In Eq.(20), the variable $m$ only appears in the last term which is just $F_e$. We know that,
\begin{widetext}
\begin{equation}
F_e^{(w)} \le \frac{1}{2} + \frac{{\sqrt {1 - r} }}{{2\sqrt {(1 - rp + {n^2}rp)[r - rp + {n^2}(1 - r + rp)} }}
\end{equation}
\end{widetext}
and the equality obtained when
\begin{equation}
m = \sqrt {\frac{{(1 - p)r + {n^2}(1 - r + pr)}}{{1 - pr + {n^2}pr}}} .
\end{equation}
Taking the relation above into the mean fidelity given in Eq.(20), we have,
\begin{widetext}
\begin{equation}
{\overline F ^{(w)}} = \frac{1}{3} + \frac{1}{6}\left( {\frac{{{n^2}(1 - r + rp)}}{{r - rp + {n^2}(1 - r + rp)}} + \frac{{1 - rp}}{{1 - rp + {n^2}rp}} + \frac{{2\sqrt {1 - r} }}{{\sqrt {(1 - rp + {n^2}rp)[r - rp + {n^2}(1 - r + rp)} }}} \right).
\end{equation}
\end{widetext}

One can find that,
\begin{equation}
{\overline F ^{(w)}} \le \frac{1}{3} + \frac{1}{6}{\left[ {1 + \frac{1}{{\sqrt {{r^2}p(1 - p)}  + \sqrt {(1 - rp)(1 - r + rp)} }}} \right]^2}.
\end{equation}
and the equality obtained when $n = \sqrt[4]{{\frac{{(1 - p)(1 - rp)}}{{p(1 - r + rp)}}}}$, Substituting the value of n into Eq.(36),we have $m = \sqrt[4]{{\frac{{(1 - p)(1 - r + pr)}}{{p(1 - pr)}}}}$.
\section*{APPENDIX C}
In this section, we give a proof that when $\Lambda_2$ has the maximal value, then Eq.(33) should be satisfied.${\Lambda _2} = \frac{{2{n_1}{n_2}(\left| E \right| - \sqrt {BC} )}}{{n_1^2n_2^2A + n_1^2B + n_2^2C + D}} \le \frac{{2{n_1}{n_2}(\left| E \right| - \sqrt {BC} )}}{{n_1^2n_2^2A + 2{n_1}{n_2}\sqrt {BC}  + D}}$,and "=" stands iff $\frac{{{n_1}}}{{{n_2}}} = \sqrt {\frac{C}{D}} $. And $\frac{{2{n_1}{n_2}(\left| E \right| - \sqrt {BC} )}}{{n_1^2n_2^2A + 2{n_1}{n_2}\sqrt {BC}  + D}} = \frac{{2(\left| E \right| - \sqrt {BC} )}}{{{n_1}{n_2}A + 2\sqrt {BC}  + \frac{D}{{{n_1}{n_2}}}}} \le \frac{{\left| E \right| - \sqrt {BC} }}{{\sqrt {BC}  + \sqrt {AD} }}$, the = stands iff ${n_1}{n_2} = \sqrt {\frac{D}{A}} $. Thus when $\Lambda_2$ has the maximal value $\frac{{\left| E \right| - \sqrt {BC} }}{{\sqrt {BC}  + \sqrt {AD} }}$, We should have Eq.(33).

\end{document}